\documentclass[prd,twocolumn,showpacs,floatfix,amsmath,nofootinbib,amssymb,floatfix]{revtex4}
\usepackage{graphicx,color,dcolumn,booktabs,bm}
\usepackage{longtable,lscape}
\usepackage{pdfpages}
\usepackage{txfonts}
\usepackage{overpic}
\usepackage{amssymb}
\usepackage{indentfirst}
\usepackage{feynmf}   
\usepackage{slashed}  
\usepackage{cases}
\usepackage{color}
\usepackage{multirow}
\usepackage{threeparttable}
\usepackage{epstopdf}
\usepackage{enumerate}
\usepackage{graphicx,color,dcolumn,booktabs,bm}
\usepackage[colorlinks, citecolor=blue,anchorcolor=red,menucolor=red, linkcolor=red,filecolor=red,urlcolor=blue,frenchlinks=red]{hyperref}

\graphicspath{{Figures/}} %

\begin{document}

\title{Newly observed $\Lambda_c(2860)^+$ at LHCb and its \emph{D}-wave partners $\Lambda_c(2880)^+$, $\Xi_c(3055)^+$ and $\Xi_c(3080)^+$}
\author{Bing Chen$^{1,3}$}\email{chenbing@shu.edu.cn}
\author{Xiang Liu$^{2,3}$\footnote{Corresponding author}}\email{xiangliu@lzu.edu.cn}
\author{Ailin Zhang$^{4}$}\email{zhangal@staff.shu.edu.cn}
\affiliation{$^1$Department of Physics, Anyang Normal University,
Anyang 455000, China\\$^2$School of Physical Science and Technology,
Lanzhou University,
Lanzhou 730000, China\\
$^3$Research Center for Hadron and CSR Physics, Lanzhou University
$\&$ Institute of Modern Physics of CAS,
Lanzhou 730000, China\\
$^4$Department of Physics, Shanghai University, Shanghai 200444,
China}

\date{\today}

\begin{abstract}
New resonance $\Lambda_c(2860)^+$, reported by LHCb, is a key to establish \emph{D}-wave charmed and charmed-strange baryon families. In this work, we first carry out an analysis of mass spectra of the $\lambda$-mode excited charmed and charmed-strange baryon states, which may reveal the relation of $\Lambda_c(2860)^+$, $\Lambda_c(2880)^+$, $\Xi_c(3055)^+$ and $\Xi_c(3080)^+$, $i.e.$, $\Lambda_c(2860)^+$ and $\Lambda_c(2880)^+$ can form a \emph{D}-wave doublet $[3/2^+,5/2^+]$ while $\Xi_c(3055)^+$ and $\Xi_c(3080)^+$ are the strange partners of $\Lambda_c(2860)^+$ and $\Lambda_c(2880)^+$, respectively. Further study of their two-body Okubo-Zweig-Iizuka-allowed decays supports $\Lambda_c(2860)^+$ and $\Xi_c(3055)^+$ as the \emph{D}-wave charmed and charmed-strange baryons, respectively, with $J^P=3/2^+$. For $\Lambda_c(2880)^+$ and $\Xi_c(3080)^+$, however, there exist some difficulties under the \emph{D}-wave assignment with $J^P=5/2^+$ since the experimental widths and some ratios of partial width of $\Lambda_c(2880)^+$ and $\Xi_c(3080)^+$ cannot be reproduced in our calculation.
\end{abstract}
\pacs{12.39.Jh,~13.30.Eg,~14.20.Lq} \maketitle

\section{Introduction}\label{sec1}

Very recently, a new charmed baryon state, which is named as $\Lambda_c(2860)^+$, was found by LHCb in the $D^0p$ channel~\cite{Aaij:2017vbw}. Its mass and decay width were determined as
\begin{equation}
\begin{split}
&m_{\Lambda_c(2860)^+}=2856.1^{+2.0}_{-1.7}\textrm{(stat)}\pm0.5\textrm{(syst)}^{+1.1}_{-5.6}\textrm{(model)~MeV},\\&
\Gamma_{\Lambda_c(2860)^+}=67.6^{+10.1}_{-8.1}\textrm{(stat)}\pm1.4\textrm{(syst)}^{+5.9}_{-20.0}\textrm{(model)~MeV}.\nonumber
\end{split}
\end{equation}
Additionally, experimental analysis indicated that it has spin-parity quantum number $J^P=3/2^+$. Before the observation of LHCb, there was possible evidence of $\Lambda_c(2860)^+$ existing in the BaBar data~\cite{Aubert:2006sp}. This newly observed $\Lambda_c(2860)^+$ directly confirms the predictions in Refs.~\cite{Chen:2014nyo,Chen:2016iyi,Lu:2016ctt,Chen:2016phw}, where a \emph{D}-wave charmed baryon around 2.85 GeV and with $J^P=3/2^+$ was suggested to be accompanied with the observed charmed baryon $\Lambda_c(2880)^+$~\cite{Olive:2016xmw}.

In the past years, more and more charmed baryons were reported with the experimental progress (see a recent review paper \cite{Chen:2016spr} for more details). Due to the experimental and theoretical effort, \emph{S}-wave and \emph{P}-wave charmed baryon families were established step by step. It is obvious that it is not the end of whole story. We are expecting  what happen to the \emph{D}-wave charmed baryon family. The LHCb's observation of $\Lambda_c(2860)^+$ is a key point to reveal it.

Firstly, we briefly introduce the experimental information of other three charmed baryons related to the present work, which are $\Lambda_c(2880)^+$, $\Xi_c(3055)^+$ and $\Xi_c(3080)^+$. $\Lambda_c(2880)^+$ was first observed by the CLEO Collaboration in the $\Lambda_c^+\pi^+\pi^-$ channel~\cite{Artuso:2000xy}, confirmed by BaBar in the $D^0p$ invariant mass spectrum~\cite{Aubert:2006sp} and Belle in the $\Sigma_c^{(\ast)}\pi$ channel~\cite{Abe:2006rz}. The available experimental analysis suggests that the $J^P=5/2^+$ assignment to $\Lambda_c(2880)^+$ is favorable. The mass and width of $\Lambda_c(2880)^+$ were measured as
\begin{equation}
\begin{split}
&M_{\Lambda_c(2880)^+}=2881.75\pm0.29\textrm{(stat)}\pm0.07\textrm{(syst)}^{+0.14}_{-0.20}\textrm{(model)},\\&
\Gamma_{\Lambda_c(2880)^+}=5.43^{+0.77}_{-0.71}\textrm{(stat)}\pm0.29\textrm{(syst)}^{+0.75}_{-0.00}\textrm{(model)},\nonumber
\end{split}
\end{equation}
in MeV by LHCb recently~\cite{Aaij:2017vbw}. In addition, the ratio
\begin{equation}
\frac{\mathcal{B}(\Lambda_c(2880)^+\rightarrow\Sigma^\ast_c(2520)\pi)}{\mathcal{B}(\Lambda_c(2880)^+\rightarrow\Sigma_c(2455)\pi)}=0.225\pm0.062\pm0.025, \label{eq1}
\end{equation}
has been given by Belle~\cite{Abe:2006rz}.

{$\Xi_c(3055)^+$ and $\Xi_c(3080)^+$ were already found in the channels of $\Lambda_c^+K^-\pi^+$, $\Sigma^{(\ast)++}_cK^-$ by Belle~\cite{Chistov:2006zj} and BaBar~\cite{Aubert:2007dt}. In last year, $\Xi_c(3055)^+$ and $\Xi_c(3080)^+$ were first observed by Belle in the $D^+\Lambda$ channel~\cite{Kato:2016hca}. With more accurate measurement, the resonance parameters of $\Xi_c(3055)^+$ obtained by Belle~\cite{Kato:2016hca} were
\begin{equation}
\begin{split}
&M_{\Xi_c(3055)^+}=3055.8\pm0.4\textrm{(stat)}\pm0.2\textrm{(syst)~MeV},\\&
\Gamma_{\Xi_c(3055)^+}=7.8\pm1.2\textrm{(stat)}\pm1.5\textrm{(syst)~MeV}.\nonumber
\end{split}
\end{equation}
Meanwhile, the mass and width of $\Xi_c(3080)^+$ were estimated to be $3077.9\pm0.9$ MeV and $3.0\pm0.7\pm0.4$ MeV, respectively. In addition, the following ratios of branching fractions
\begin{equation}
\frac{\mathcal{B}(\Xi_c(3055)^+\rightarrow\Lambda D^+)}{\mathcal{B}(\Xi_c(3055)^+\rightarrow\Sigma_c(2455)^{++}K^-)}=5.09\pm1.01\pm0.76, \label{eq2}
\end{equation}
\begin{equation}
\frac{\mathcal{B}(\Xi_c(3080)^+\rightarrow\Lambda D^+)}{\mathcal{B}(\Xi_c(3080)^+\rightarrow\Sigma_c(2455)^{++}K^-)}=1.29\pm0.30\pm0.15, \label{eq3}
\end{equation}
and
\begin{equation}
\frac{\mathcal{B}(\Xi_c(3080)^+\rightarrow\Sigma_c(2520)^{++}K^-)}{\mathcal{B}(\Xi_c(3080)^+\rightarrow\Sigma_c(2455)^{++}K^-)}=1.07\pm0.27\pm0.01, \label{eq4}
\end{equation}
were also reported~\cite{Kato:2016hca}, where the uncertainties are statistical and systematic.

It is obvious that the information above is crucial to find the relation among these four states and identify their properties. In this work, we carry out a mass spectrum analysis by combing with these observed  $\Lambda_c(2860)^+$, $\Lambda_c(2880)^+$, $\Xi_c(3055)^+$, and $\Xi_c(3080)^+$, which may suggest that these four states are 1\emph{D} candidates in charmed baryon family. For further testing this assignment to them, we need to perform the study of two-body Okubo-Zweig-Iizuka (OZI) allowed strong decay behavior of them, where the Eichten-Hill-Quigg (EHQ) decay formula is adopted.

This paper is organized as follows. After introduction, we give a mass spectrum analysis of these discussed $\Lambda_c(2860)^+$, $\Lambda_c(2880)^+$, $\Xi_c(3055)^+$, and $\Xi_c(3080)^+$ in Sec. \ref{sec2}. In Sec. \ref{sec3}, the two-body OZI-allowed decays of these four charmed baryons will be discussed. The paper ends with the discussion and conclusion in Sec. \ref{sec4}.

\section{The mass spectrum analysis of \emph{D}-wave charmed and charmed-strange baryons}\label{sec2}

\begin{table}[b]
\caption{Comparison of the theoretical results with experimental data for the $\lambda$-mode excited 1\emph{D} $\Lambda_c$ and $\Xi_c$ baryon masses (in MeV).} \label{table1}
\renewcommand\arraystretch{1.2}
\begin{tabular*}{85mm}{@{\extracolsep{\fill}}lcccc}
\toprule[1pt]\toprule[1pt]
Assignments &\multicolumn{2}{c}{$1D(3/2^+)$}  & \multicolumn{2}{c}{$1D(5/2^+)$}  \\
\cline{2-3}\cline{4-5}
 Candidates       & $\Lambda_c(2860)^+$  & $\Xi_c(3055)^+$& $\Lambda_c(2880)^+$  & $\Xi_c(3080)^+$ \\
\midrule[0.8pt]
 Expt.~\cite{Aaij:2017vbw,Kato:2016hca}    & 2856.1    & 3055.8   & 2881.8     & 3077.9      \\
 Ref.~\cite{Chen:2014nyo}    & 2857    & 3055   & 2879     & 3076      \\
 Ref.~\cite{Chen:2016iyi}    & 2843    & 3033   & 2851     & 3040      \\
 Ref.~\cite{Ebert:2011kk}    & 2874    & 3059   & 2880     & 3076    \\
 Ref.~\cite{Roberts:2007ni}  & 2887    & 3012   & 2887     & 3004        \\
 Ref.~\cite{Shah:2016mig}    & 2873    & 3080   & 2849     & 3054        \\
\bottomrule[1pt]\bottomrule[1pt]
\end{tabular*}
\end{table}

Many works have focused on the mass spectra of higher excited charmed baryons, where different phenomenological models were adopted which include the relativistic flux tube (RFT) model~\cite{Chen:2014nyo}, the potential models~\cite{Chen:2016iyi,Lu:2016ctt,Ebert:2011kk,Roberts:2007ni,Shah:2016mig}, the QCD sum rule~\cite{Chen:2016phw}, and the Regge phenomenology~\cite{Guo:2008he}. In Table~\ref{table1}, we collect some predicted masses for \emph{D}-wave charmed and charmed-strange baryons. In Refs.~\cite{Chen:2014nyo,Ebert:2011kk,Chen:2016iyi,Roberts:2007ni,Shah:2016mig}, the $\lambda$-mode excitations of charmed baryons were investigated, where
the orbital excitation only exists between light quark cluster (two light quarks) and charm quark. Just shown in Table \ref{table1}, the mass of the 1\emph{D} charmed baryon with $J^P=3/2^+$ is about $2840\sim 2890$ MeV. Thus, the newly observed $\Lambda_c(2860)^+$ can be assigned as the \emph{D}-wave $3/2^+$ charmed baryon with $\lambda$ mode excitation well. As expected by theories (see Table \ref{table1}), a $5/2^+$ $\Lambda_c^+$ should be accompanied with $\Lambda_c(2860)^+$ in the near mass region.
Therefore, $\Lambda_c(2880)^+$ becomes to be a good $5/2^+$ candidate of \emph{D}-wave charmed baryon with $\lambda$ mode excitation (see the comparison between experimental and theoretical results in Table \ref{table1}). In a word, we may specify that $\Lambda_c(2860)^+$ with $J^P=3/2^+$ and $\Lambda_c(2880)^+$ with $J^P=5/2^+$ form a degenerate \emph{D}-wave doublet $[3/2^+,5/2^+]$ in the heavy quark limit, which can be reflected by the small mass difference between $\Lambda_c(2860)^+$ and $\Lambda_c(2880)^+$.

\begin{figure}[t]
\begin{center}
\includegraphics[width=8.4cm,keepaspectratio]{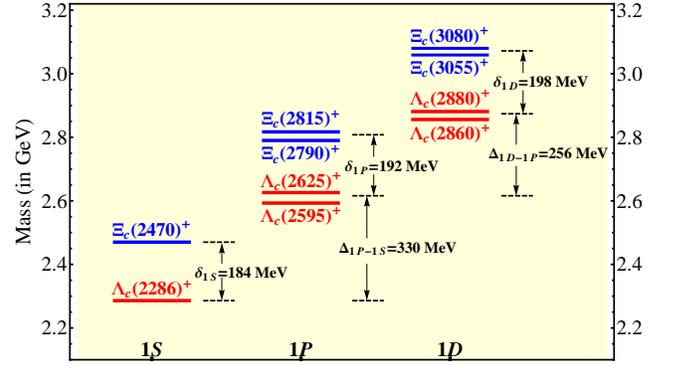}
\caption{The experimentally observed $\Lambda_c^+$ and $\Xi_c^+$ states. The \emph{S}-wave ($\Lambda_c(2286)^+$ and $\Xi_c(2470)^+$) and \emph{P}-wave ($\Lambda_c(2595)^+$, $\Lambda_c(2625)^+$, $\Xi_c(2790)^+$, and $\Xi_c(2815)^+$) members have been well established~\cite{Olive:2016xmw}, while the \emph{D}-wave state candidates ($\Lambda_c(2860)^+$, $\Lambda_c(2880)^+$, $\Xi_c(3055)^+$, and $\Xi_c(3080)^+$) are still in dispute. Here, $\delta_{1i}$ ($i=S,\,P,\,D$) denotes the mass gaps between \emph{i}-wave $\Lambda_c$ and $\Xi_c$ states. $\Delta_{1P-1S}$ refers to the mass difference between the \emph{P}-wave and \emph{S}-wave excited states, while $\Delta_{1D-1P}$ denotes the mass gap between the \emph{D}-wave and \emph{P}-wave excited states.}\label{Fig1}
\end{center}
\end{figure}

When checking the masses of $\Lambda_c(2860)^+$, {$\Lambda_c(2880)^+$, $\Xi_c(3055)^+$ and $\Xi_c(3080)^+$}, we find the mass relations
\begin{eqnarray}
&&M_{\Xi_c(3080)^+}-M_{\Lambda_c(2880)^+}\approx M_{\Xi_c(3055)^+}-M_{\Lambda_c(2860)^+},\\
&&M_{\Lambda_c(2880)^+}-M_{\Lambda_c(2860)^+}\approx M_{\Xi_c(3080)^+}-M_{\Xi_c(3055)^+},
\end{eqnarray}
which show that the observed $\Xi_c(3055)^+$ and $\Xi_c(3080)^+$ can be strange partners of $\Lambda_c(2860)^+$ and $\Lambda_c(2880)^+$, respectively. More specifically, $\Xi_c(3055)^+$ and $\Xi_c(3080)^+$ can be grouped into a doublet $[3/2^+,5/2^+]$ in \emph{D}-wave $\Xi_c$ family (as shown in Fig. \ref{Fig1}). This assignment to $\Xi_c(3055)^+$ and $\Xi_c(3080)^+$ can be also supported by theoretical results since the 1\emph{D} $\Xi_c$ states with $J^P=3/2^+$ and $J^P=5/2^+$ were predicted with mass about  $3000\sim 3090$ MeV~\cite{Chen:2014nyo,Ebert:2011kk,Chen:2016iyi,Roberts:2007ni,Shah:2016mig} (see Table \ref{table1} for more details).

If we consider the SU(3) flavor symmetry, the comparable mass splittings in 1\emph{S} and 1\emph{P} charmed and charmed-strange baryons~\cite{Olive:2016xmw} can be explained by a RFT model~\cite{Chen:2014nyo}. And then, the mass splitting (22 MeV) for the $\lambda$-mode excited \emph{D}-wave $3/2^+$ and $5/2^+$ $\Lambda_c$ states was predicted in Ref.~\cite{Chen:2014nyo}, which is in good agreement with the experimental data (26 MeV). In addition, the averaged mass of $\Lambda_c(2860)^+$ and $\Lambda_c(2880)^+$ is about 256 MeV higher than that of the 1\emph{P} states, which is also consistent with the theoretical estimate in Ref. \cite{Chen:2014nyo}.

When further comparing the masses of well-established \emph{S}-wave and \emph{P}-wave $\Lambda_c/\Xi_c$ states~\cite{Olive:2016xmw}, we also find that the mass gaps between the corresponding $\Lambda_c$ and $\Xi_c$ states is about 190 MeV (see Fig.\ref{Fig1}). This law for the mass difference between $\Lambda_c$ and $\Xi_c$ families has also been explained by the RFT model~\cite{Chen:2014nyo}, by which the  strange partners of $\Lambda_c(2860)^+$ and $\Lambda_c(2880)^+$ are predicted around 3060 MeV. This estimate provides an extra support of the observed $\Xi_c(3055)^+$~\cite{Aubert:2007dt,Kato:2016hca} and $\Xi_c(3080)^+$~\cite{Chistov:2006zj,Aubert:2007dt,Kato:2016hca} as the strange partners of $\Lambda_c(2860)^+$ and $\Lambda_c(2880)^+$, respectively.

By the above analysis of mass spectrum, we may conclude that four observed $\Lambda_c(2860)^+$, $\Lambda_c(2880)^+$, $\Xi_c(3055)^+$ and $\Xi_c(3080)^+$ are categorized into $\lambda$-mode excitations of \emph{D}-wave charmed and charmed-strange baryons. To test this assignment to them, in following, we will perform a detailed study of their two-body OZI-allowed strong decays, by which their partial and total decay widths can be obtained.

\section{The decay behavior of these discussed states}\label{sec3}

In this section, we will employ the EHQ decay formula which was proposed by Eichten, Hill, and Quigg (EHQ)~\cite{Eichten:1993ub} to calculate the two-body OZI-allowed decays of $\Lambda_c(2860)^+$, $\Lambda_c(2880)^+$, $\Xi_c(3055)^+$ and $\Xi_c(3080)^+$. The general expression of the EHQ decay formula reads
\begin{equation}\label{eq7}
\Gamma^{A\rightarrow BC}_{j_C,\ell} = \xi\,\left|\mathcal
{C}^{s_Q,j_B,J_B}_{j_C,j_A,J_A}\mathcal
{M}^{j_A,j_B}_{j_C,\ell}(q)\right|^2 \,q^{2\ell+1}\,
e^{-q^2/\tilde{\beta}^2}.
\end{equation}
Here, the flavor factors, $\xi$, have been given in Ref.~\cite{Chen:2016iyi}. $q=|\vec q|$ denotes the three-momentum of a final state in the rest frame of an initial state. \emph{A} and $B$ represent the initial and final heavy-light hadrons, respectively. {The total angular momenta and the corresponding projections are defined as $J_i$ and $j_i$ ($i=A,B$), respectively. Here
$C$ denotes the light flavor hadron. The explicit expression of $\tilde{\beta}$ has been given in our previous work~\cite{Chen:2016iyi} (see Appendix A in Ref. \cite{Chen:2016iyi} for more details). In addition, the normalized coefficient $\mathcal {C}^{s_Q,j_B,J_B}_{j_C,j_A,J_A}$ is given by the following equation
\begin{eqnarray}\label{eq8}
\begin{split}
\mathcal {C}^{s_Q,j_B,J_B}_{j_C,j_A,J_A}=(-1)^{J_A+j_B+j_C+s_Q}~&\sqrt{(2j_A+1)(2J_B+1)}\\ &\times\left\{
           \begin{array}{ccc}
                    s_Q  & j_B & J_B\\
                    j_C  & J_A & j_A\\
                    \end{array}
     \right\},
\end{split}
\end{eqnarray}
where $\vec{j}_C \equiv \vec{s}_C + \vec{\ell}$. The symbols $s_C$ and $\ell$ represent the spin of the light hadron $C$ and the orbital angular momentum relative to $B$, respectively. {The spin of heavy quark $Q$ is denoted as $s_Q$.} The coefficient given by Eq.~(\ref{eq8}) has explicitly incorporated into {the heavy quark symmetry} which is believed to be important for the strong decays of heavy-light hadrons~\cite{Isgur:1991wq}.

The transition factors $\mathcal {M}^{j_A,j_B}_{j_C,\ell}(q)$ which is related to the non-perturbative dynamics can be calculated by various phenomenological models. {In this work, the transition factors can be calculated by the $^3P_0$ model~\cite{Micu:1968mk,LeYaouanc:1972vsx,LeYaouanc:1988fx}. In the following, we briefly introduce the procedure of calculating these transition factors by the $^3P_0$ model. Firstly, the mock states of initial and final hadrons are constructed, where a simple harmonic oscillator (SHO) wave function is applied to describe the spatial wave function of a hadron state involved in these discussed decays. And then, the helicity amplitude $\mathcal{M}^{j_A,j_B,j_C}(q)$ can be deduced by the transition operator $\hat{\mathcal{T}}$ of the $^3P_0$ model. With the help of Jacob-Wick formula, the partial wave amplitudes $\mathcal{M}_{LS}(q)$ are related to
the obtained helicity amplitude $\mathcal{M}^{j_A,j_B,j_C}(q)$.  By performing a unitary rotation between the $LS$ coupling and $jj$ coupling, finally, the transition factors $\mathcal {M}^{j_A,j_B}_{j_C,\ell}(q)$ can be extracted directly. More details for this approach can be found in Ref.~\cite{Chen:2016iyi}.} For different decay modes of the \emph{D}-wave charmed baryons, the explicit expressions for $\mathcal {C}^{s_Q,j_B,J_B}_{j_C,j_A,J_A}$ and $\mathcal {M}^{j_A,j_B}_{j_C,\ell}(q)$ are listed in Table \ref{table2}. With the above preparation, the partial and total widths can be calculated for the \emph{D}-wave charmed baryons.

\begin{table}[t]
\caption{The transition factors and the normalized coefficients for the decays of \emph{D}-wave $\Lambda_c$ and $\Xi_c$ states appeared in the EHQ formula. $\mathfrak{B}^\prime_c(1S)$ and $\mathfrak{B}^\ast_c(1S)$ denote the $1/2^+$ and $3/2^+$ $\Sigma_c/\Xi^\prime_c$ states, respectively. $\mathcal {P}$ is the light pseudoscalar mesons while ``LB'' denotes the light baryons in final states. $\mathfrak{B}^{\prime0,1/2}_c(1P)$ is abbreviation of $\Sigma_c(2700)$ and $\Xi^\prime_c(2840)$.} \label{table2}
\renewcommand\arraystretch{1.5}
\begin{tabular*}{85mm}{@{\extracolsep{\fill}}ccrcc}
\toprule[1pt]\toprule[1pt]
$~J^P$  & $\mathfrak{B}^\prime_c(1S)$~+~$\mathcal {P}$  & $\mathfrak{B}^\ast_c(1S)$~+~$\mathcal {P}$  & $\mathfrak{B}^{\prime0,1/2}_c(1P)$~+~$\mathcal {P}$ &  \emph{D}~+~``LB'' \\
\midrule[0.8pt]
 $~\frac{3}{2}^+$  & $\sqrt{\frac{5}{6}}\mathcal {M}^{2,1}_{1,1}(q)$ & $\sqrt{\frac{1}{6}}\mathcal {M}^{2,1}_{1,1}(q)$ & $\mathcal {M}^{2,0}_{2,2}(q)$ & $\sqrt{\frac{5}{8}}\mathcal {M}^{2,1/2}_{3/2,1}(q)$   \\
                      &        & $\mathcal {M}^{2,1}_{3,3}(q)$  &        &      \\
 $~\frac{5}{2}^+$  &        & $\mathcal {M}^{2,1}_{1,1}(q)$ & $\mathcal {M}^{2,0}_{2,2}(q)$      &      \\
                      & $-\frac{\sqrt{5}}{3}\mathcal {M}^{2,1}_{3,3}(q)$   & $\frac{2}{3}\mathcal {M}^{2,1}_{3,3}(q)$ &      &   $-\sqrt{\frac{5}{12}}\mathcal {M}^{2,1/2}_{5/2,3}(q)$   \\
\bottomrule[1pt]\bottomrule[1pt]
\end{tabular*}
\end{table}

\subsection{$\Lambda_c(2860)^+$ and $\Lambda_c(2880)^+$}

\begin{table}[b]
\caption{The partial and total decay widths in MeV, and branching fractions in \%, of $\lambda$-mode excited \emph{D}-wave $\Lambda_c$ states.} \label{table3}
\renewcommand\arraystretch{1.2}
\begin{tabular*}{85mm}{@{\extracolsep{\fill}}lcccc}
\toprule[1pt]\toprule[1pt]
Decay  &\multicolumn{2}{c}{$\Lambda_c(2860)^+~[1D(3/2^+)]$}  & \multicolumn{2}{c}{$\Lambda_c(2880)^+~[1D(5/2^+)]$}  \\
\cline{2-3}\cline{4-5}
modes  & $\Gamma_i$  & $\mathcal{B}_i$  & $\Gamma_i$  & $\mathcal{B}_i$ \\
\midrule[0.8pt]
 $\Sigma_c(2455)\pi$       & 2.2     & 3.0\%    & 0.4       & 1.7\%      \\
 $\Sigma^\ast_c(2520)\pi$  & 1.0     & 1.4\%    & 3.7       & 15.4\%    \\
 $\Sigma_c(2700)\pi$       & 0.0     & 0.0\%    & 0.2       & 0.8\%      \\
 $D^0p$                    & 34.5    & 48.0\%   & 10.8      & 44.8\%        \\
 $D^+n$                    & 34.2    & 47.6\%   & 9.0       & 37.3\%        \\
 \midrule[0.8pt]
  Theory                   & 71.9    & 100\%    & 24.1      & 100\%      \\
  Expt.~\cite{Aaij:2017vbw} & \multicolumn{2}{l}{$67.6^{+10.1}_{-8.1}\pm1.4^{+5.9}_{-20.0}$}  &   \multicolumn{2}{l}{$5.43^{+0.77}_{-0.71}\pm0.29^{+0.75}_{-0.00}$}    \\
\bottomrule[1pt]\bottomrule[1pt]
\end{tabular*}
\end{table}

If treating $\Lambda_c(2860)^+$ as a $\lambda$-mode excited \emph{D}-wave state with $J^P=3/2^+$, the obtained partial and total decay widths of $\Lambda_c(2860)^+$ are presented in Table~\ref{table3}. The result shows that the $D^0p$ and $D^+n$ channels are the main decay modes for $\Lambda_c(2860)^+$, which can explain why $\Lambda_c(2860)^+$ was firstly observed in the $D^0p$ channel. We notice that there is no evidence of $\Lambda_c(2860)^+$ observed in the $\Sigma_c(2455)\pi$ and $\Sigma^\ast_c(2520)\pi$ channels, in which the $\Lambda_c(2880)^+$ has been detected by Belle~\cite{Abe:2006rz}. Our result can reflect this fact since the partial widths of the $\Lambda_c(2860)^+$ decays into $\Sigma_c(2455)\pi$ and $\Sigma^\ast_c(2520)\pi$ are quite small. Additionally, the calculated total decay width of $\Lambda_c(2860)^+$ is 71.9 MeV which is in agreement with the LHCb data well~\cite{Aaij:2017vbw}. Therefore, we may conclude that {\it $\Lambda_c(2860)^+$ as the $\lambda$-mode excited D-wave state with $J^P=3/2^+$ can be established well}.

In the following, we continue to discuss $\Lambda_c(2880)^+$. This state has been detected in the $D^0p$ channel by BaBar~\cite{Aubert:2006sp} and LHCb~\cite{Aaij:2017vbw}. These measurements indicate that $D^0p$ is a primary decay channel of $\Lambda_c(2880)^+$. If $\Lambda_c(2880)^+$ is the $\lambda$-mode excited \emph{D}-wave state with $J^P=5/2^+$, the branching ratio of its $D^0p$ channel can reach up to 44.8\%, which can explain why $\Lambda_c(2880)^+$ was observed in the $D^0p$ channel. {The $D^+n$ channel is its main decay modes. Besides the $D^0 p$ and $D^+ n$ as dominant decay channels, we notice the decay channel $\Sigma^\ast_c(2520)\pi$ has considerable contribution to its total width. }
However, we find some difficulties to assign $\Lambda_c(2880)^+$ as a \emph{D}-wave state with $J^P=5/2^+$. Firstly, the calculated total decay width is 4.4 times larger than the experimental width of $\Lambda_c(2880)^+$. However, the measurement given by CLEO~\cite{Artuso:2000xy} and Belle~\cite{Abe:2006rz} indicates that $\Sigma^\ast_c(2520)\pi$ is a less important decay mode for $\Lambda_c(2880)^+$. Thirdly, we cannot reproduce the ratio listed in Eq. (\ref{eq1}). Thus, if categorizing $\Lambda_c(2880)^+$ as the $\lambda$-mode excited \emph{D}-wave state with $J^P=5/2^+$, we have to be faced with these three contradictions indicated above, and should provide reasonable solution. In Sec. \ref{sec4}, we will discuss this point.

\subsection{$\Xi_c(3055)^+$ and $\Xi_c(3080)^+$}

We study the decay properties of $\Xi_c(3055)^+$ as the $\lambda$-mode excited \emph{D}-wave charmed-strange baryon with $J^P=3/2^+$. The results presented in Table \ref{table4} show that $D^+\Lambda$ and $\Sigma_c(2455)K$ are its dominant decay modes, which naturally explain why $\Xi_c(3055)^+$ was firstly found in the $\Sigma_c(2455)^{++}K^-$ channel by BaBar~\cite{Aubert:2007dt}, and confirmed by Belle in the $\Sigma_c(2455)^{++}K^-$ and $D^+\Lambda$ channels~\cite{Kato:2016hca}.
In addition, the theoretical value for the $\Gamma(D^+\Lambda)/\Gamma(\Sigma_c(2455)K)$ ratio is about 2.3 which is comparable with the experimental data (see Eq. (\ref{eq2})) if considering the experimental error.

The obtained total width of $\Xi_c(3055)^+$ as a \emph{D}-wave charmed-strange baryon with $J^P=3/2^+$ is about 15.4 MeV, which is comparable with the upper limit of Belle's result ($5.1\sim10.5$ MeV). Based on the analysis of mass spectrum and the study of the decay behavior, $\Xi_c(3055)^+$ could be regarded as the $3/2^+$ \emph{D}-wave state. In other word, $\Xi_c(3055)^+$ may be the strange partner of $\Lambda_c(2860)^+$. This assignment to $\Xi_c(3055)^+$ is also supported by an analysis in the chiral quark model~\cite{Liu:2012sj}.
Under this assignment, we have given a strong prediction of the spin-parity quantum number of $\Xi_c(3055)^+$, $i.e.$, $\Xi_c(3055)^+$ has $J^P=3/2^+$. This tough prediction can be tested directly by further experiment study.

\begin{table}[htbp]
\caption{The partial and total decay widths in MeV, and branching fractions in \%, of the $\lambda$-mode excited \emph{D}-wave $\Xi_c$ states.} \label{table4}
\renewcommand\arraystretch{1.2}
\begin{tabular*}{85mm}{@{\extracolsep{\fill}}lcccc}
\toprule[1pt]\toprule[1pt]
Decay  &\multicolumn{2}{c}{$\Xi_c(3055)~[1D(3/2^+)]$}  & \multicolumn{2}{c}{$\Xi_c(3080)~[1D(5/2^+)]$}  \\
\cline{2-3}\cline{4-5}
modes  & $\Gamma_i$  & $\mathcal{B}_i$  & $\Gamma_i$  & $\mathcal{B}_i$ \\
\midrule[0.8pt]
 $\Sigma_c(2455)K$        & 4.2     & 27.3\%   & 0.6       & 4.8\%      \\
 $\Xi^\prime_c(2580)\pi$  & 0.2     & 1.3\%    & 0.1       & 0.8\%      \\
 $\Sigma^\ast_c(2520)K$   & 0.7     & 4.5\%    & 5.4       & 42.8\%    \\
 $\Xi^\ast_c(2645)\pi$    & 0.3     & 2.0\%    & 0.5       & 4.0\%      \\
 $\Xi^\prime_c(2840)\pi$  & 0.4     & 2.6\%    & 0.7       & 5.5\%      \\
 $D^+\Lambda$             & 9.6     & 62.3\%   & 5.3       & 42.1\%        \\
 \midrule[0.8pt]
  Theory                  & 15.4    & 100\%    & 12.6      & 100\%      \\
  Expt.~\cite{Kato:2016hca} & \multicolumn{2}{l}{$7.8\pm1.2\pm1.5$}  &   \multicolumn{2}{l}{$3.0\pm0.7\pm0.4$}    \\
\bottomrule[1pt]\bottomrule[1pt]
\end{tabular*}
\end{table}

As a $\lambda$-mode excited \emph{D}-wave charmed-strange baryon with $J^P=5/2^+$, $\Xi_c(3080)^+$ has two main decay modes $\Sigma_c(2520)K$ and $D^+\Lambda$. We notice that $\Xi_c(3080)^+$ was observed in the channels of $\Lambda_c^+K^-\pi^+$~\cite{Chistov:2006zj}, $\Sigma^{(\ast)++}_cK^-$~\cite{Aubert:2007dt,Kato:2016hca}, and $D^+\Lambda$~\cite{Kato:2016hca}. Although our theoretical result can reasonably explain why $\Xi_c(3080)^+$ was found in the $\Sigma^{\ast++}_cK^-$~\cite{Aubert:2007dt,Kato:2016hca}, and $D^+\Lambda$~\cite{Kato:2016hca} channels, we cannot reproduce the ratios listed in Eqs.~(\ref{eq2})-(\ref{eq3}) since our calculation shows that the partial decay width of the $\Sigma_c(2455)K$ channel is only about 0.6 MeV. So it is far smaller than that of $\Sigma_c(2520)K$ (see Table~\ref{table4}). Similar to the situation of $\Lambda_c(2880)^+$, the obtained total decay width of $\Xi_c(3080)^+$ is about 4 times larger than experimental measurement~\cite{Kato:2016hca}. Then we also need to face the contradictions above if we assign $\Xi_c(3080)^+$ as a $\lambda$-mode excited \emph{D}-wave charmed-strange baryon with $J^P=5/2^+$.

{Before closing this section, we need to discuss the theoretical uncertainties of the results presented in Tables \ref{table3} and \ref{table4}.
The main uncertainties may come from the approximations used to obtain the EHQ decay formula, adopting a simple harmonic oscillator wave function to replace the hadron wave functions, and ignoring relativistic effect.
Additionally, we adopt experimental widths like $\Sigma_c(2520)^{++}\rightarrow\Lambda_c(2286)^+\pi^+$~\cite{Chen:2016iyi}
to fix the value of parameter $\gamma$ and take the experimental masses of initial and final states as input, which provide extra sources of uncertainty of these obtained result. 
}

\section{Discussion and conclusion}\label{sec4}

Stimulated by newly reported $\Lambda_c(2860)^+$ \cite{Aaij:2017vbw}, we carry a comprehensive study of $\Lambda_c(2860)^+$ associated with three observed resonances $\Lambda_c(2880)^+$, $\Xi_c(3055)^+$, and $\Xi_c(3080)^+$. The mass spectrum analysis shows that
$\Lambda_c(2860)^+$, $\Lambda_c(2880)^+$, $\Xi_c(3055)^+$, and $\Xi_c(3080)^+$ can be grouped into the $\lambda$-mode excited \emph{D}-wave charmed or charmed-strange baryon families. $\Lambda_c(2860)^+$ with $\Lambda_c(2880)^+$ or $\Xi_c(3055)^+$ with $\Xi_c(3080)^+$ may form a degenerate \emph{D}-wave doublet of charmed or charmed-strange baryon, which  also indicates that $\Xi_c(3055)^+$ and $\Xi_c(3080)^+$ are strange partners of $\Lambda_c(2860)^+$ and $\Lambda_c(2880)^+$, respectively.

For further testing this assignment, we carry out a study of their two-body OZI-allowed strong decays, which provide valuable information of their partial and total decay widths.
For $\Lambda_c(2860)^+$, the assignment as a $\lambda$-mode excited \emph{D}-wave charmed baryon with $J^P=3/2^+$ is suggested since the obtained theoretical results of partial and total decay widths can reproduce the experimental data well. And then, $\Xi_c(3055)^+$ as the strange partner of $\Lambda_c(2860)^+$ is also supported by the study of its decay behavior. In fact, the present study gives an strong constraint of the spin-parity quantum number of $\Xi_c(3055)^+$, where $\Xi_c(3055)^+$ has $J^P=3/2^+$. This prediction is a crucial information to further test the \emph{D}-wave assignment to $\Xi_c(3055)^+$ in future experiment.

Although a spectrum analysis strongly suggests that $\Lambda_c(2880)^+$ and $\Xi_c(3080)^+$ can be explained as the $\lambda$-mode excited \emph{D}-wave charmed and charmed-strange baryons, respectively, with $J^P=5/2^+$. However, there exist some difficulties when we further study their strong decay properties, $i.e.$, the total decay widths of $\Lambda_c(2880)^+$ and $\Xi_c(3080)^+$ obtained in our model is far larger than the experimental widths. And some experimental ratios of the partial decays cannot be reproduced in the present scenario.

{If $\Lambda_c(2880)^+$ and $\Xi_c(3080)^+$ are difficult to be identified as $\lambda$-mode excited \emph{D}-wave charmed and charmed-strange baryons, respectively, we need to further search for other possible assignments to $\Lambda_c(2880)^+$ and $\Xi_c(3080)^+$. For example, $\Lambda_c(2880)^+$ and $\Xi_c(3080)^+$ as $\rho$-mode excited states or $\rho$-mode and $\lambda$-mode simultaneously excited states in the \emph{D}-wave charmed and charmed-strange baryon families should be tested by the mass spectrum analysis and the study of their decay behavior. Besides, the mixing effect of the different \emph{D}-wave states with $J^P=5/2^+$ should be considered in future work.
In addition, we notice the $D^*N$ and $DN$ molecular state assignments to the observed $\Lambda_c(2940)^+$ and $\Sigma_c(2800)$ in literatures \cite{He:2006is,Dong:2010gu}. However, for the $\Lambda_c(2880)^+$ and $\Xi_c(3080)^+$, the hadronic molecular state assignments are not suitable since molecular states are usually predicted to have masses a few MeV below meson-baryon thresholds with corresponding quantum numbers. Whereas there are no such thresholds in the vicinity of the $5/2^+$ states, $\Lambda_c(2880)^+$ and $\Xi_c(3080)^+$.
Considering the present situation of $\Lambda_c(2880)^+$ and $\Xi_c(3080)^+$, we suggest further experimental study of the resonance parameters of $\Lambda_c(2880)^+$ and $\Xi_c(3080)^+$ and the ratios of partial widths. Of course, more theoretical studies of $\Lambda_c(2880)^+$ and $\Xi_c(3080)^+$ by different phenomenological models are encouraged.

In summary, in this work we identify the possibility of the observed $\Lambda_c(2860)^+$, $\Lambda_c(2880)^+$, $\Xi_c(3055)^+$, and $\Xi_c(3080)^+$ as the \emph{D}-wave charmed and charmed-strange baryons. We have reasons to believe that \emph{D}-wave charmed and charmed-strange baryon families will become more and more abundant with experimental progress and theoretical effort. Finally, the \emph{D}-wave $\Lambda_c$ and $\Xi_c$ states will be established, which is an interesting research issue in the coming years.

\section*{Acknowledgement}

This project is supported by the National Natural Science Foundation of China under Grant Nos. 11305003, 11222547, 11175073, 11447604, 11647301 and 11475111. Xiang Liu is also supported by the National  Program for Support of Top-notch Young Professionals and the Fundamental Research Funds for the Central Universities.


\end{document}